\documentclass[aip,twocolumn,a4paper,amsmath,floatfix]{revtex4}
\usepackage[utf8]{inputenc}
\usepackage{graphicx}
\usepackage{amsmath}
\usepackage{amssymb}
\usepackage{bm}

\usepackage{color}

\begin{document}
\title{Fast event-driven simulations for soft spheres: from dynamics to Laves phase nucleation}
\author{Antoine Castagn\`ede$^{1,}$}
\email[\textbf{Author to whom correspondence should be addressed: }]{antoine.castagnede@universite-paris-saclay.fr}
\affiliation{$^1$Universit\'e Paris-Saclay, CNRS, Laboratoire de Physique des Solides, 91405 Orsay, France}
\author{Laura Filion$^2$}
\affiliation{$^2$Soft Condensed Matter and Biophysics, Debye Institute for Nanomaterials Science, Utrecht University, Utrecht, Netherlands}
\author{Frank Smallenburg$^1$}
\affiliation{$^1$Universit\'e Paris-Saclay, CNRS, Laboratoire de Physique des Solides, 91405 Orsay, France}

\begin{abstract}
Conventional molecular dynamics (MD) simulations struggle when simulating particles with steeply varying interaction potentials, due to the need to use a very short time step. Here, we demonstrate that an event-driven Monte Carlo (EDMC) approach first introduced by Peters and de With [Phys. Rev. E \textbf{85}, 026703 (2012)] represents an excellent substitute for MD in the canonical ensemble. In addition to correctly reproducing the static thermodynamic properties of the system, the EDMC method closely mimics the dynamics of systems of particles interacting via the steeply repulsive Weeks-Chandler-Andersen (WCA) potential. In comparison to time-driven MD simulations, EDMC runs faster by over an order of magnitude at sufficiently low temperatures. Moreover, the lack of a finite time step in EDMC circumvents the need to trade accuracy against simulation speed associated with the choice of time step in MD. We showcase the usefulness of this model to explore the phase behavior of the WCA model at extremely low temperatures, and to demonstrate that spontaneous nucleation and growth of the Laves phases is possible at temperatures significantly lower than previously reported. 
\end{abstract}\maketitle

\section{Introduction}
\label{section:introduction}

Conventional molecular dynamics (MD) simulations function by numerically integrating Newton's equations of motion for all particles in the system. To do this, MD simulations typically discretize time using a fixed time step. Ideally, this time step is chosen as large as possible to optimize the overall simulation speed. However, to obtain accurate results, the time step must be chosen small enough to avoid systematic errors due to the approximations involved in the discretization of time. Generally, the time step should be small enough that the forces on any given particle do not change significantly during a single step. This poses a significant challenge for many-body systems where the interaction potential and its derivatives are sharply varying functions of the interparticle distance.

The extreme version of an isotropic steeply varying interaction potential is the hard-sphere model. In this fundamental model system \cite{royall2023colloidal}, the interaction potential between two particles jumps from zero to infinity at contact, rendering conventional MD algorithms useless. Intriguingly, this problem can be overcome via a different strategy: event-driven molecular dynamics (EDMD) simulations \cite{alder1959studies, rapaport2009event, bannerman2011dynamo,de2011optimizing, smallenburg2022efficient}. In the event-driven approach, collisions between particles are resolved as instantaneous events, which can be efficiently predicted based on the observation that between collisions, hard particles move in simple straight lines at constant velocity. However, this approach usually limits event-driven methods to systems of particles where the interaction potential only changes via discontinuous jumps, such as hard spheres and square-well or square-shoulder models (see e.g. \cite{rapaport2009event,  bannerman2009structure, bannerman2011dynamo,dotera2014mosaic, akkaya2015event, paganini2015structure, plati2024quasi}).

In principle, it would be interesting to have simulation methods available that benefit from the potential efficiency of event-driven approaches but are still capable of handling particles with continuously varying interaction potentials.
One way of accomplishing this is the discretization of continuous potentials \cite{thomson2014mapping, ucyigitler2012optimization}, although this necessarily results in approximated interactions. A novel alternative route was introduced in 2012 by Peters and de With \cite{peters2012rejection}, who developed an event-driven approach to simulate particles with continuous interaction potentials in the canonical ensemble. In this approach, which can be regarded as a rejection-free Monte Carlo method, the particles move in straight lines between discrete collisions with their neighbors. The collision distance between two particles is stochastically chosen based on the Boltzmann distribution, ensuring that the method correctly samples the canonical ensemble. By handling the interactions between pairs of particles as discrete events, this method neatly sidesteps the issues associated with finite integration time steps encountered in conventional MD of steep potentials. The approach of Peters and de With has proven useful in the extension of event-chain Monte Carlo (ECMC) methods \cite{bernard2009event} to continuous potentials \cite{peters2012rejection, bernard2012addendum, michel2014generalized}. ECMC simulations are an extremely efficient method to sample the static properties of many-body systems (especially those involving long-range interactions \cite{kapfer2016cell}), made possible by the use of simplified, unphysical dynamics involving chains of colliding particles \cite{bernard2009event}. However, this approach makes them less suitable for studying the dynamics of these systems. In contrast, the approach originally proposed in Ref. \cite{peters2012rejection} is capable of modeling systems with near-physical dynamics, where all particles move collectively. This raises an interesting question: can this event-driven Monte Carlo (EDMC) approach be efficiently used as a substitute for molecular dynamics when exploring the dynamics of particles with almost-hard interactions?

Here, we combine the approach of Peters and de With with an efficient implementation of event-driven molecular dynamics, constructing an efficient simulation tool for exploring the behavior of short-ranged, steeply varying potentials. We test this method on systems of steeply repulsive particles (modeled via the Weeks-Chandler-Andersen (WCA) potential \cite{weeks1971role}), in order to assess its similarity to Newtonian dynamics and its efficiency. As expected, EDMC correctly reproduces the static equilibrium properties of the simulated systems, such as the equation of state and its phase behavior. More importantly, over a large temperature regime, the EDMC \textit{dynamics} also closely resemble MD trajectories in terms of diffusion and structural relaxation. Moreover, EDMC can be simulated significantly more efficiently, especially close to the hard-sphere limit where conventional MD struggles. To highlight the usefulness of this method, we show two applications based on recent literature studies of low-temperature WCA models: one investigating the low-temperature phase behavior of the WCA model \cite{attia2022comparing}, and one examining the effect of softness on the nucleation of the Laves phases in almost-hard spheres \cite{dasgupta2020tuning}.

In the remainder of this manuscript, we first describe the EDMC method and its application to the WCA potential. Next, we apply it to monodisperse systems of WCA particles and examine its dynamics and phase behavior. Finally, we use it to examine nucleation in binary mixtures of WCA particles of two different sizes. We end with a brief discussion and conclusions.

\section{Methods}
\label{section:methods}

\subsection{Event-driven Monte Carlo simulations}

In Ref. \onlinecite{peters2012rejection}, Peters and de With proposed an innovative event-driven approach to simulate many-body systems with continuous pair potentials.
Their approach is a rejection-free Monte Carlo method in the canonical ensemble (i.e. at fixed number of particles $N$, volume $V$, and temperature $T$). Similar to an EDMD simulation, the particles move in straight lines between collisions, and collide as if they are hard spheres. However, the distance of collision is stochastically determined based on the interaction potential and temperature of the system.
As a result, the traditional acceptance rule of the Metropolis scheme \cite{frenkel2023understanding} is replaced with a stochastic choice of a maximum energy gain between two particles before a collision occurs. In particular, the collision distance $r_\mathrm{col}$ between two particles is chosen such that the total energy gain of the particles as they approach is equal to a random energy $\Delta U$ taken from the Boltzmann distribution at the chosen temperature $T$. Any decrease in energy during this interval is ignored. At the time of collision, the two particles exchange momentum as if they are two hard spheres undergoing an elastic collision. As shown in Ref. \onlinecite{peters2012rejection}, it can be proven that this correctly reproduces the equilibrium distribution in the canonical ensemble.

\begin{figure}
    \centering
    \includegraphics[width=0.48\textwidth]{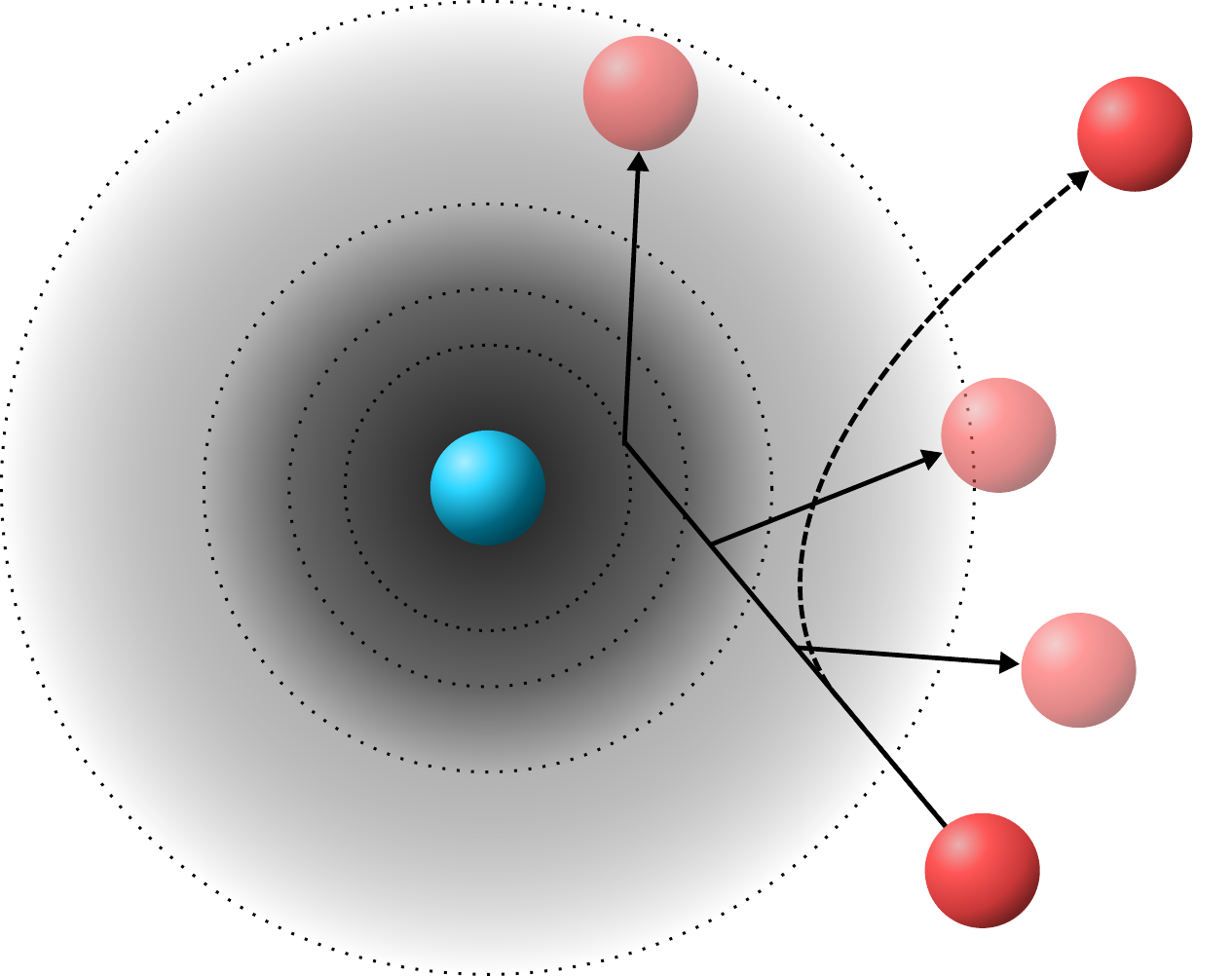}
    \caption{Schematic view of the collision process between two particles as processed by the conventional time-driven MD approach (dashed line), and the present EDMC approach (solid lines). The stochastic nature of the EDMC simulation method can yield a range of different trajectories. Note that for simplicity, the blue particle is considered to be fixed. The background shading and dotted contour lines illustrate the interaction potential $U(r)$, with darker shading indicating higher energy.
    }
    \label{figure:EDMCschematic}
\end{figure}

If we compare the dynamics generated by this EDMC scheme to those expected in a standard MD simulation (in the micro-canonical ensemble), then it is clear that the dynamics of this rejection-free method are not identical for any pair potential. Firstly, EDMC natively simulates the canonical ensemble, while MD simulations would require a thermostat to do so. Moreover, as illustrated in Fig. \ref{figure:EDMCschematic}, in a normal MD trajectory of repulsive spheres, the velocities of a pair of ``colliding'' particles change continuously as the particles approach and separate again. In contrast, the EDMC trajectories consist of straight lines, meeting at a sharp corner. Although these two trajectories converge in the limit of infinitely steep repulsive (i.e. hard-sphere) potentials, this implies that the microscopic dynamics of the two simulation methods can in principle be expected to differ significantly. The effect of this change on the macroscopic dynamics remains to be investigated.

Event-driven simulations can be implemented extremely efficiently for isotropic hard particles \cite{smallenburg2022efficient, klement2019efficient}, suggesting that an event-driven approach to simulate soft potentials could potentially be more efficient than time-driven schemes. This is particularly the case in the limit of hard-sphere-like potentials, where MD typically requires small integration time steps. This limit is also the regime where we would expect the dynamics resulting from EDMC and MD to be most similar, and hence where this approach may be suitable for studying dynamical features of the system under consideration.

To explore how EDMC simulations compare to MD methods both in terms of resulting dynamics and efficiency, here we apply this method to a commonly used interaction potential for sharply repulsive soft spheres: the WCA potential \cite{weeks1971role}. In this model, the pair potential between particles $i$ and $j$ is given by
\begin{equation}
    U_{ij}(r) = 
    \begin{cases}
    4\epsilon \left[ 
    \left(\dfrac{\sigma_{ij}}{r}\right)^{12} -  
    \left(\dfrac{\sigma_{ij}}{r}\right)^{6}
    + \dfrac{1}{4}\right], & \dfrac{r}{\sigma_{ij}}
    < 2^{1/6} \\
    0 & \mathrm{otherwise}        
    \end{cases},\label{eq:WCApot}
\end{equation}
where $r$ is the distance between two particles, $\epsilon$ controls the strength of the interaction, and $\sigma_{ij} = (\sigma_i + \sigma_j)/2$ with $\sigma_i$ the diameter of particle $i$.
The cutoff for this WCA potential can then be written as $\sigma_\mathrm{cut} = 2^{1/6} \sigma_{ij}$. Note that in the low-temperature limit, this system reduces to a hard-particle model, with the contact distance given by $\sigma_\mathrm{HS} = \sigma_\mathrm{cut}$.
Note that Eq. \ref{eq:WCApot} can be analytically inverted to obtain the distance corresponding to a certain pair energy $U$:
\begin{equation}
  r_\mathrm{WCA}(U) = \left(\frac{2}{1 + \sqrt{U / \epsilon}}\right)^{1/6} \sigma_{ij}. 
\end{equation}

To simulate the WCA potential, we modify an efficient implementation of EDMD for hard spheres \cite{smallenburg2022efficient}. The main change is in the prediction of collisions. For two spheres $i$ and $j$ that move in straight lines and collide at a separation distance $r_\mathrm{col}$, the time of collision can be analytically predicted based on the instantaneous positions and velocities of said particles:
\begin{equation}
    t_\mathrm{col} = \frac{ -\mathbf{r}_{ij} \cdot \mathbf{v}_{ij} -\sqrt{\left(\mathbf{r}_{ij} \cdot \mathbf{v}_{ij}\right)^2 - v_{ij}^2 \left(r_{ij}^2 -r_\mathrm{col}^2\right)}}{v_{ij}^2},
    \label{eq:collisiontime}
\end{equation}
where $\mathbf{r}_{ij} = \mathbf{r}_j - \mathbf{r}_i$ is the vector indicating the relative position of the two particles, and $\mathbf{v}_{ij} = \mathbf{v}_j - \mathbf{v}_i$ is their relative velocity. For hard spheres, $r_\mathrm{col}$ is simply equal to $\sigma_{ij}$. For the WCA potential, we instead find the collision distance stochastically, using \cite{peters2012rejection}
\begin{equation}
    r_\mathrm{col} = r_\mathrm{WCA}(U_{ij}(t) + \Delta U),
\end{equation}
where $U_{ij}(t)$ is the pair energy between the two particles at the current simulation time, and $\Delta U$ is a random energy drawn from the (exponential) Boltzmann distribution at the chosen temperature $T$ (see SM for implementation details \cite{blackman2021scrambled, ahrens1972computer,devroye1986continuous}). Equation \ref{eq:collisiontime} then provides the moment of collision. If the argument of the square root is negative, no solution exists, and the two particles will not collide. Note that since the WCA potential is purely repulsive, we only have to consider pairs of particles that are currently approaching each other ($\mathbf{r}_{ij} \cdot \mathbf{v}_{ij} < 0$).

Predicted collisions are resolved in the event-driven fashion as if they are normal hard-sphere collisions, conserving kinetic energy and momentum. Note that as a result, this scheme inherently conserves the total kinetic energy of the system, while the total potential energy fluctuates. Hence, it is important to ensure that at the start of the simulation, the average kinetic energy of the particles matches the chosen temperature $T$.

\subsection{Comparing EDMC and MD}

For comparison, we also simulate our systems using the simulation package LAMMPS \cite{LAMMPS}, using a Nos\'e-Hoover thermostat to keep the temperature fixed.

Simulation times are compared in terms of the ``thermal'' time unit
\begin{equation}
 \tau_{kT} = \sqrt{\beta m \sigma^2},
\end{equation}
where $\beta = 1/k_B T$ with $k_B$ Boltzmann's constant, $m$ is the mass of a particle (which we always choose equal for all particles), and our length unit $\sigma$ is chosen to be the diameter of the largest sphere size in our system. This time unit is a measure for the time it takes a free particle at temperature $T$ to move a distance on the order of $\sigma$. Additionally, it is convenient to introduce an ``energetic'' time unit 
\begin{equation}
\tau_\epsilon = \sqrt{\frac{m \sigma^2}{\epsilon}} = \sqrt{\frac{k_B T}{\epsilon}} \tau_{kT},
\end{equation}
as this is more conventionally used in time-driven simulations.

To initialize our systems in the fluid state, we make use of the hard-sphere EDMD code of Ref. \onlinecite{smallenburg2022efficient}, where the particles are grown as purely hard spheres from a random dilute initial configuration to reach the desired packing fraction.
These configurations are then equilibrated with the WCA potential at the temperature of interest, using either EDMC or LAMMPS. The equilibrated configurations are used as initial configurations for the simulations where we perform our measurements of e.g. the potential energy, pressure, diffusion coefficient, intermediate scattering function, and simulation speed. 
For the specific case of benchmarking, we simulate multiple runs over a time of $100 \tau_\textrm{kT}$, each sequentially starting from the last snapshot of the previous one. 
Benchmarking results were obtained on a dedicated machine running Ubuntu 20.04 and powered by two 8-core 3.3Ghz Intel Xeon Gold 6234 processors. All simulations used only a single core.

\section{Model}

In this paper, we consider both monodisperse and bidisperse systems of WCA particles. 

\subsection{Monodisperse WCA}

The phase behavior of the monodisperse WCA model has been studied extensively (see e.g. Refs. \onlinecite{ahmed2009phase, attia2022comparing}), and consists of a fluid at low densities and a face-centered cubic crystal at high densities, separated by a coexistence region. In the limit of low temperatures ($k_B T / \epsilon \to 0$), the WCA model reduces to hard spheres with an effective diameter equal to $2^{1/6}\sigma$. As a result, the freezing and melting densities in this limit can be determined by simply rescaling the hard-sphere binodals \cite{smallenburg2024simple}, and are given by $\rho_F \sigma^3 \simeq 0.664$ and $\rho_M \sigma^3 = 0.774$, respectively. At higher temperatures, both the freezing and melting densities shift to higher values, as the particles effectively shrink. 

As such, it is common to map the freezing transition of the monodisperse WCA system onto that of a pure hard sphere (HS) system. This way, one can define a temperature-dependent effective diameter $\sigma_\textrm{eff}$ for the WCA particles and thus a temperature-dependent effective reduced density $\rho^* = \rho \sigma_\textrm{eff}^3$ for the system. To this end, we use the fit of the WCA freezing density $\rho_F$ introduced by Filion \textit{et al.} \cite{filion2011simulation} based on the data from Ahmed and Sadus \cite{ahmed2009phase}:
\begin{equation}
    \rho_F\sigma^3 \simeq 0.635 + 0.473 \left(\dfrac{k_B T}{\epsilon}\right)^{1/2} - 0.236\, \dfrac{k_B T}{\epsilon}.
    \label{eq:effHSfit}
\end{equation}

\subsection{Binary WCA}

As our second model, we consider a binary WCA mixture, with size ratio $\gamma = \sigma_s/\sigma_L = 0.78$ and fixed composition $x_L = N_L / \left(N_L + N_S \right) = 1/3$. Here, the subscripts $S,L$ denote the small and large particles, respectively. In the low-temperature limit, this model reduces to a binary hard-sphere mixture of the same size ratio. Hard-sphere mixtures of approximately this size ratio are known to be capable of forming a stable binary crystal phase with the same structure as MgZn$_2$, one of the Laves phases \cite{hynninen2009stability}. In fact, the other two Laves phase structures (MgCu$_2$ and MgNi$_2$) are only slightly less stable than MgZn$_2$ for  hard-sphere mixtures as well, with only small free-energy differences separating them \cite{hynninen2009stability}.  

The low-temperature behavior of this WCA mixture, and in particular the interplay between glassy dynamics and the nucleation of the Laves crystal phases, has been studied previously by Dasgupta, Coli, and Dijkstra in Ref. \onlinecite{dasgupta2020tuning}. They observed the spontaneous nucleation of the Laves phases at relatively high temperatures $k_B T/\epsilon \gtrsim 0.025$, but concluded that at low temperatures, where the model approaches the hard-sphere limit, spontaneous nucleation is pre-empted by dynamical arrest associated with the glass transition. In particular, they determine both an instability line, marking the state points where the fluid phase becomes unstable with respect to crystallization, and a glass transition line, where the fluid is no longer able to dynamically relax, and argued that these cross at a low temperature on the order of $k_BT/\epsilon \simeq 0.01$. However, we note that this observation conflicts with recent observations of spontaneous nucleation of Laves phases in simulation studies of binary mixtures of true hard spheres \cite{bommineni2020spontaneous, marin2020tetrahedrality}, making this an interesting question to revisit.

To detect nucleation in the binary WCA model, we perform local structure identification in each snapshot using a supervised machine learning algorithm. In particular, we use a single-layer neural network classifier, following the methodology from Ref. \onlinecite{boattini2018neural}. The network is trained to recognize particles whose local environment matches one of the different considered phases, specifically fluid, FCC, and each of the three Laves phases (MgCu$_2$, MgZn$_2$, and MgNi$_2$).
The performed analysis is based on characterization of the local environment of particles in the system using the locally averaged bond order parameters introduced by Lechner and Dellago \cite{lechner2008accurate}. 
The weights and biases of the machine learning model we use in our work have been obtained by training this model on a set of snapshots of fluid, FCC, and all three Laves phases generated from simulations of a binary hard-sphere mixture with the same size ratio and composition as our chosen binary WCA mixture. Note that this system corresponds to the zero-temperature limit of our binary WCA model. We report that the accuracy of the model on the training set exceeds $98\%$ for all considered phases, similar to the performances reported in Ref. \onlinecite{boattini2018neural}.

\section{Test case 1: Monodisperse WCA model}
\label{section:monoWCA}

We first focus on the monodisperse WCA model.
In order to explore the dynamics produced by the EDMC method, we initially focus on a dense fluid phase at density $\rho\sigma^3 = 0.65$, which is just below freezing in the zero-temperature limit, and remains in the fluid phase at higher temperatures. At temperatures above $k_B T / \epsilon \simeq 0.01$, this density is in the stable fluid regime instead \cite{ahmed2009phase, attia2022comparing}.

\begin{figure*}
    \raggedright
    a) \hspace{155pt} b) \hspace{155pt} c)\\
    \centering
    \includegraphics[width=0.325\textwidth]{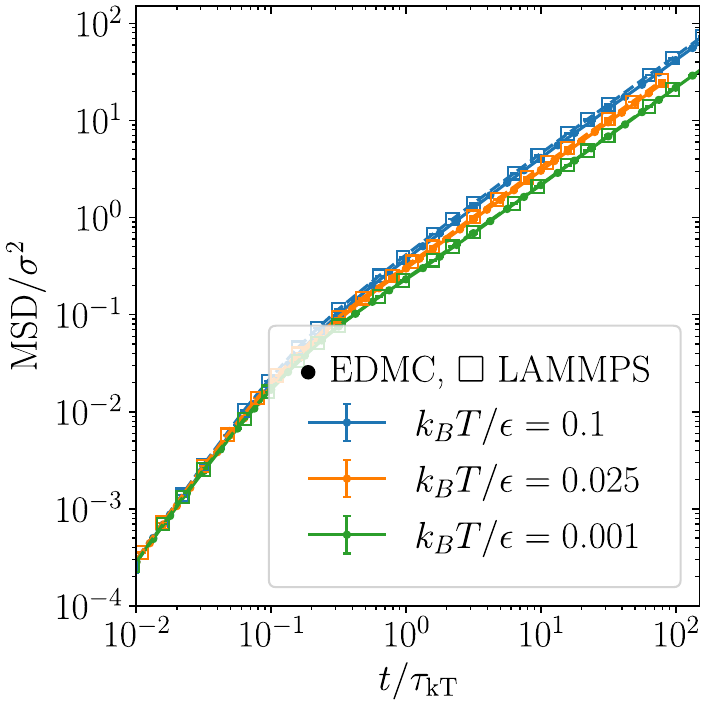} \includegraphics[width=0.325\textwidth]{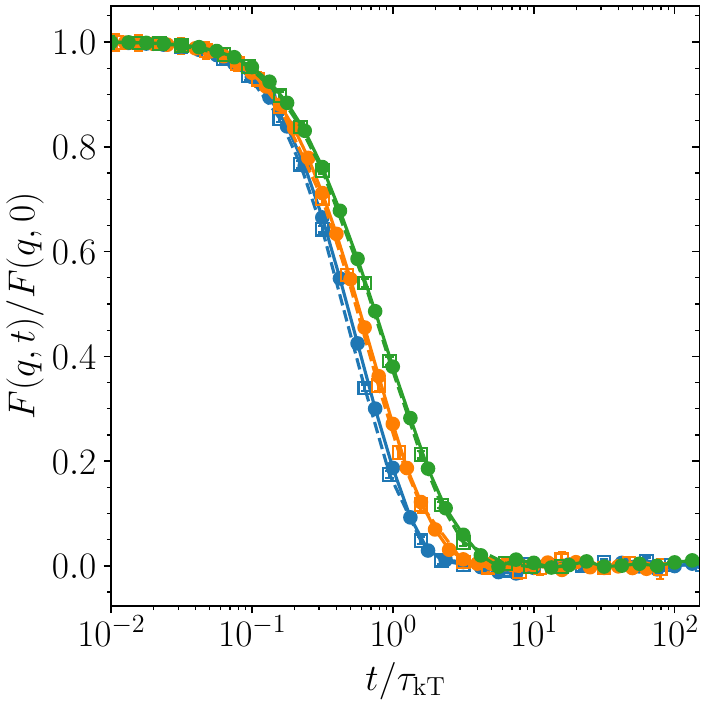} \includegraphics[width=0.325\textwidth]{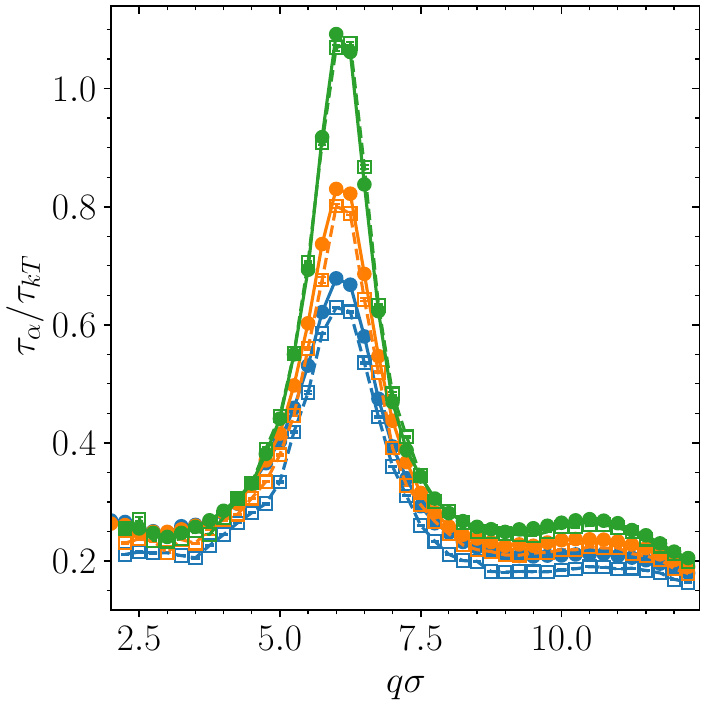}
    \caption{Comparison between dynamics of EDMC (marked by solid dots ($\bullet$) and lines) and LAMMPS (marked by empty squares ($\Box$) and dashed lines) for monodisperse WCA at density $\rho \sigma^3 = 0.65$, for a range of temperatures. a) Mean squared displacement as a function of time. b) Intermediate scattering function $F(q,t)$ ($q = 2\pi / \sigma$) as a function of time. c) Relaxation time $\tau_\alpha$ as a function of the wave-vector $q$. 
    Note that error bars (determined as twice the standard error) are shown for all points but are typically smaller than the point size. All lines are guides to the eye.
    }
    \label{figure:monoWCAdynamics}
\end{figure*}

In Fig. \ref{figure:monoWCAdynamics}a, we show the mean squared displacement of the WCA model as measured using both EDMC and LAMMPS as a function of time. We see that for all investigated temperatures, there is very little difference between the two methods. A very close examination shows mild deviations at high temperatures, with the LAMMPS simulations resulting in slightly faster dynamics than EDMC.

In Fig. \ref{figure:monoWCAdynamics}b, we examine the intermediate scattering function (ISF) $F(q,t)$ of the same systems, defined as
\begin{equation}
    F(q,t) = \frac{1}{N} \langle \rho(q,t) \rho(-q, 0) \rangle,
\end{equation}
where $\rho(q,t) = \sum_j \exp(-i \mathbf{q}\cdot \mathbf{r}_j)$ is the Fourier transform of the density and $F(q,t)$ describes the correlation of collective density fluctuations in the system.
The ISF is measured at the wave vector $q = 2\pi / \sigma$. We observe much the same results as for the mean squared displacement: EDMC simulations result in slightly slower dynamics at high temperatures, but the differences are very subtle. Finally, to confirm that the switch to EDMC does not introduce deviations in the dynamics at other length scales, we show in Fig. \ref{figure:monoWCAdynamics}c the variation of the structural relaxation time $\tau_\alpha$ (determined as the point where $F(q,t) = 1/e$) with the wave vector $q$. We see the same qualitative behavior at all examined temperatures over the range of wave vectors considered, meaning that similar relaxation dynamics are observed at all investigated length scales.

\begin{figure}
    \raggedright
    a)\\
    \centering
    \includegraphics[width=\columnwidth]{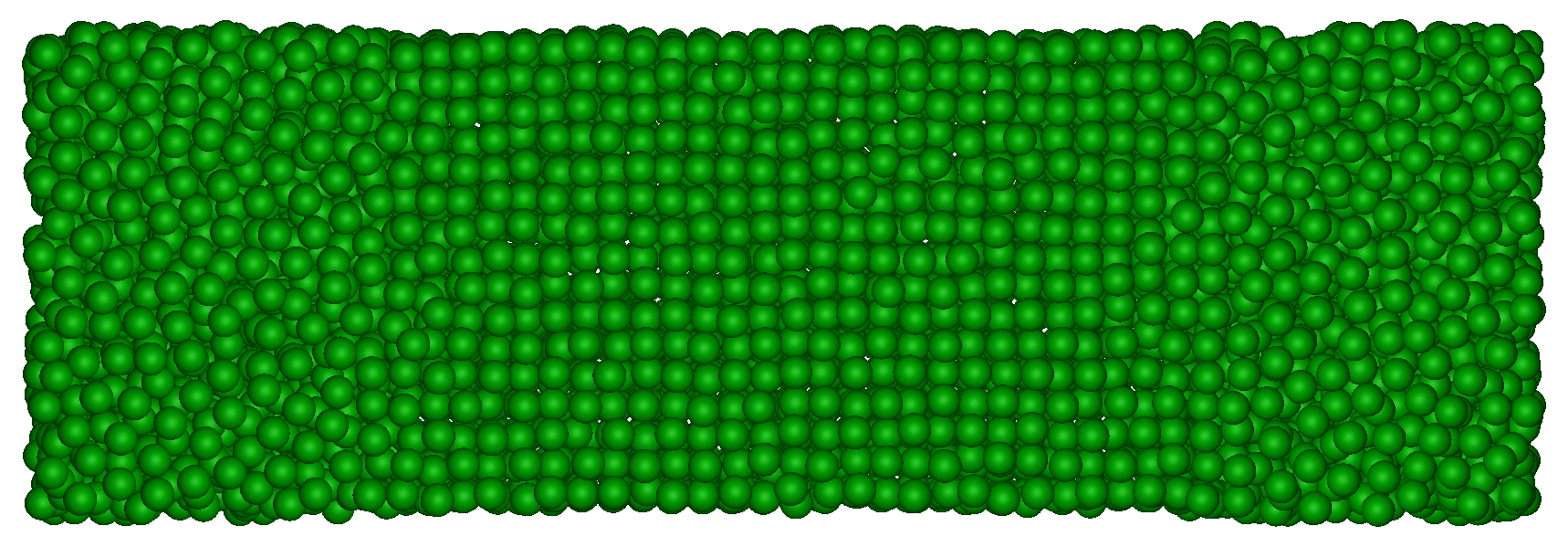}\\
    \raggedright
    b)\\
    \centering
    \includegraphics[width=\columnwidth]{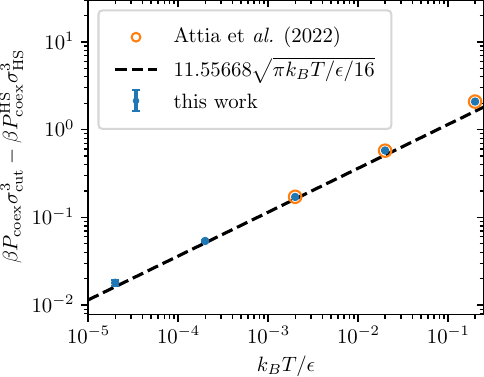}
    \caption{a) Typical snapshot of a direct coexistence simulation. b)
    WCA phase diagram at very low temperatures. Empty circles represent recent estimates also obtained from interface pinning simulations\cite{attia2022comparing}. The estimate for $\beta P_\mathrm{coex}^\mathrm{HS} \sigma_\mathrm{HS}^3= 11.55668$ was obtained using the same methodology for a pure hard-sphere system of the same size \cite{smallenburg2024simple}. Error bars are determined as twice the standard error on five independent measurements, and are mostly smaller than the points. The dashed line shows the expected low-temperature scaling (see Ref. \onlinecite{attia2022comparing}). 
    }
    \label{figure:monoWCApd}
\end{figure}

To demonstrate the effectiveness of EDMC for establishing equilibrium properties of systems near the hard-sphere limit, we turn our attention to determining the phase boundaries of the low-temperature WCA system, using the direct coexistence approach of Ref. \onlinecite{smallenburg2024simple}. In particular, we perform direct coexistence simulations in the canonical ensemble using an elongated simulation box, such that the fluid and FCC crystal phases coexist in a slab geometry, separated by two interfaces normal to the long $z$-axis of the box (see Fig. \ref{figure:monoWCApd}a). As the lattice spacing is fixed by the initial configuration and the box size in the $xy$-plane, we perform multiple runs at each temperature with several different lattice spacings chosen roughly near the melting density, and identify the lattice spacing which corresponds to the equilibrium phase coexistence (see Ref. \onlinecite{smallenburg2024simple} for more details). Figure \ref{figure:monoWCApd} shows the coexistence pressure as a function of temperature for systems of $N=6144$ particles. We compare this result to the data from Attia, Dyre, and Pedersen, who used an interface pinning approach combined with MD simulations \cite{attia2022comparing} for approximately the same system size, finding excellent agreement. A key advantage of the EDMC method is that it can readily be extended to arbitrarily low temperatures, as it does not rely on discrete time steps. Hence, we are able to extend our results to much lower temperatures than were easily accessible in the MD simulations of Ref. \onlinecite{attia2022comparing}. As expected, at low temperatures the coexistence pressure closely approaches that of a monodisperse hard-sphere system with effective particle diameter $\sigma_{HS} = 2^{1/6}\sigma$.

The above results clearly show that the EDMC approach accurately reproduces the thermodynamics and dynamics of the studied monodisperse WCA system, even at temperatures where the interaction potential is significantly removed from hard spheres. The next question that we would like to answer is how efficient this method is in comparison to a conventional MD simulation. To do this, however, we first have to establish a reasonable choice of integration time step for the MD simulations. Specifically, we want the time step $\delta t$ to be small enough to reproduce the thermodynamics of the system associated with the limit of infinitely small time steps. On the other hand, from an efficiency point of view, we want the time step to be large, such that fewer integration steps are needed to simulate the desired time interval.

\begin{figure}
    \centering
    \includegraphics[width=\columnwidth]{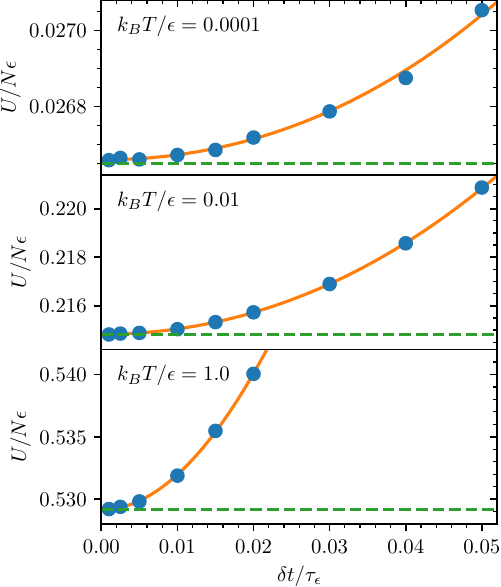} 
    \caption{MD measurements of the average energy of the WCA model at density $\rho \sigma^3 = 0.68$ as a function of the MD time step, for different temperatures as indicated. The solid orange lines are cubic fits to the data at $\delta t/\tau_\epsilon < 0.03$, and the dashed green lines indicate the extrapolation of that fit to $\delta t = 0$. For these results, we used $N = 8000$.
    }
    \label{figure:monoWCAtimestep}
\end{figure}

In Fig. \ref{figure:monoWCAtimestep}, we plot the behavior of the potential energy of the WCA system as measured in LAMMPS for different choices of the time step and for several different temperatures. Clearly, when the time step is too large, (e.g. $\delta t = 0.01 \tau_\epsilon$), we observe a shift to noticeably higher energies in comparison to the true value reached in the limit of small time steps. Based on these results, and on the weak effect of temperature on this behavior, we choose $\delta t = 0.001 \tau_\epsilon$ for all of our MD simulations in this paper. Note that the time required to perform an MD simulation over a fixed total time interval trivially scales as $(\delta t)^{-1}$. The following benchmarking results should be regarded in light of this performance dependence: by raising the time step by e.g. a factor of 10, the performance of the MD simulations could be boosted by a similar factor, at the cost of a small but noticeable deviation of the thermodynamic properties of the simulated system (see Fig. \ref{figure:monoWCAtimestep}).

\begin{figure}
    \raggedright
    a)\\
    \centering
    \includegraphics[width=\columnwidth]{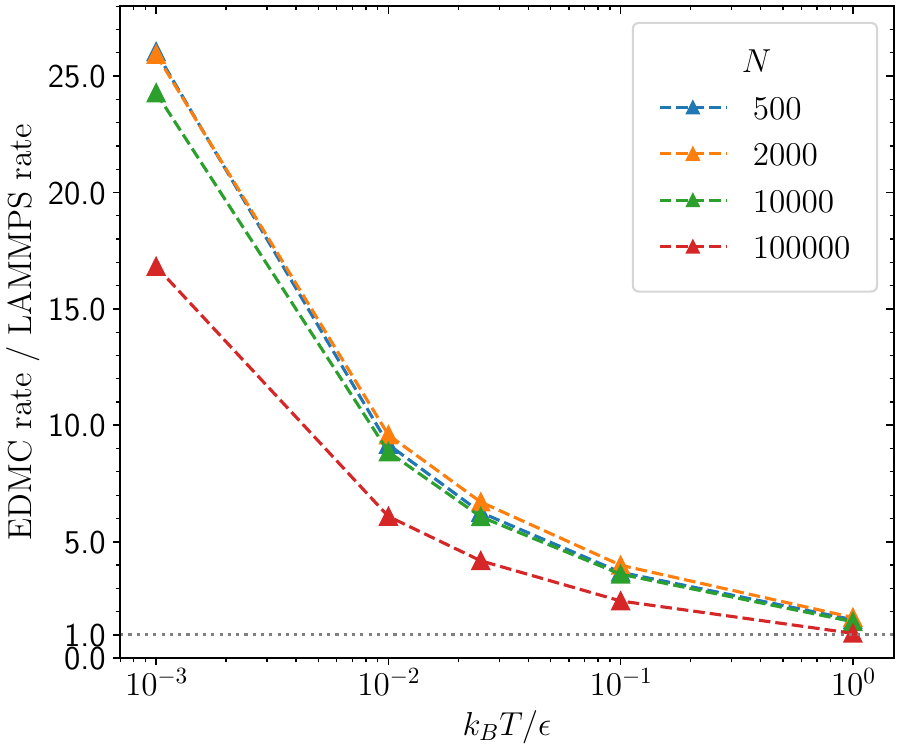}\\
    \raggedright
    b)\\
    \centering
    \includegraphics[width=\columnwidth]{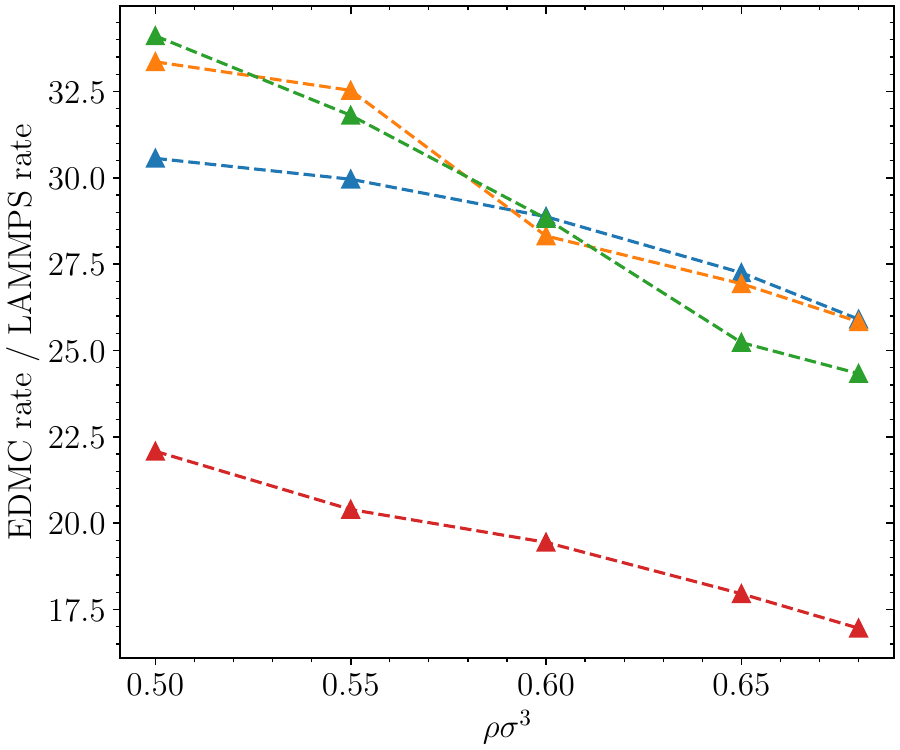}
    \caption{Relative performance of the EDMC and MD methods for monodisperse WCA systems of several  sizes, expressed as the ratio of the number of time units simulated per second in each simulation method, a) at fixed density $\rho \sigma^3 = 0.68$ and b) at fixed temperature $k_B T / \epsilon = 0.001$. Performance evaluation was obtained from successive short simulations of duration $100 \tau_{kT}$ started from an equilibrated fluid system. Standard deviations are typically found to be much less than $1 \%$.
    }
    \label{figure:monoWCAbench}
\end{figure}

In Fig. \ref{figure:monoWCAbench}, we show benchmarking results comparing EDMC to the the LAMMPS simulations. In particular, we plot the ratio of number of time units simulated per second in both an EDMC and a LAMMPS simulation at the same state point. In Fig. \ref{figure:monoWCAbench}a we plot this quantity for a fixed density $\rho\sigma^3 = 0.68$ as a function of $k_B T / \epsilon$, reported for a series of system sizes. We have confirmed that for all these simulations, the systems remained in the fluid state. The EDMC approach significantly outperforms the MD approach over most of the investigated temperature range, and especially in the low-temperature regime, where it is faster than MD by a factor of approximately 20. This speed-up can be understood by considering the number of simulation ``steps'' (i.e. collisions or integration steps) required to simulate $1\,\tau_\epsilon$ in time. For MD, this is fixed, as it is purely determined by the integration time step. For EDMC, this is simply the number of particle collisions during the chosen time interval, which scales $\propto \sqrt{k_B T/\epsilon}$ due to the thermal slowdown of the particles, and hence vanishes in the low-temperature limit. Note that this observation only holds because the pressure $P$ (in units of $\epsilon/\sigma^3$) in the low-temperature limit vanishes for this system: for a system in which the particles still feel interactions with their neighbors in the ground state, we would not expect the same scaling. We also observe that LAMMPS scales slightly better with system size, as can be seen from the decrease in relative performance for larger system sizes. The exact effect of system size on the relative performance is not monotonic, likely due to details in e.g. the way the number of cells in the cell list is chosen in each simulation, which can lead to discrete `steps' in performance. Although it is possible to further optimize both simulation methods by fine-tuning the neighbor list parameters for each density, we have not done so here. However, we have checked for specific state points that changing the neighbor list shell size does not significantly enhance the speed of either simulation method: for both methods this optimization leads to only small performance differences that do not exceed $\sim15\%$.

Figure \ref{figure:monoWCAbench}b shows the relative performance of the two methods as a function of density $\rho \sigma^3$ for a monodisperse WCA system equilibrated in the fluid phase at temperature $k_B T / \epsilon = 0.001$.
We observe that the relative efficiency of our simulation method slightly decreases as a function of density but consistently outperforms LAMMPS by a significant factor. The high efficiency of the EDMC simulation at low densities is understandable, since the computational effort required by this method is concentrated in handling collisions. Overall, both methods are more efficient at low densities, due to the fact that during either collision predictions or force calculations, fewer neighboring particles have to be considered. In addition to this, the EDMC simulation benefits from a lower number of collisions per unit time, while the MD simulation still needs to consider the same number of integration steps per time unit.

In short, for the case of monodisperse WCA systems, we find that the EDMC approach significantly outperforms the MD approach for a large range of (low) temperatures and system sizes. The EDMC method is competitive with MD even at temperatures around $k_B T/\epsilon = 1$, and -- as expected -- performs best in the limit of low temperatures. To examine how this method performs in a more complex system, we explore in the next section the application of the EDMC approach to a binary WCA system, and in particular examine glassy behavior and crystal nucleation.

\section{Test case 2: Laves phase nucleation in a binary WCA model}
\label{section:laves}

\begin{figure}
    \raggedright
    a)\\
    \centering
    \includegraphics[width=\columnwidth]{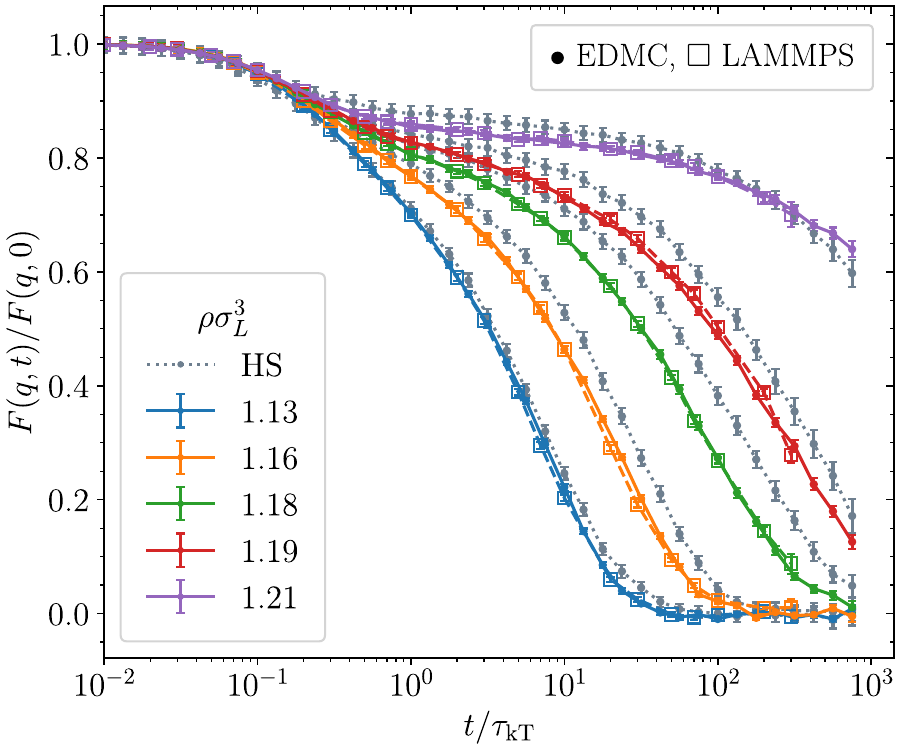}\\
    \raggedright
    b)\\
    \centering
    \includegraphics[width=\columnwidth]{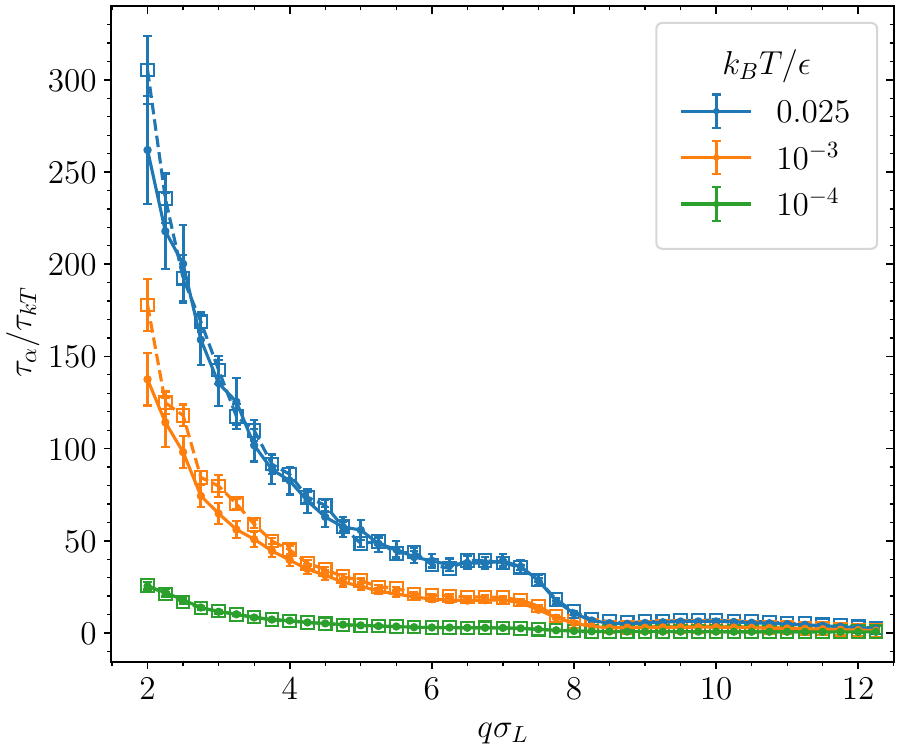}
    \caption{Comparison between dynamics of EDMC and LAMMPS for a binary WCA mixture with composition $x_L = 1/3$ and size ratio $\gamma = 0.78$. Symbols and lines are the same as in figure \ref{figure:monoWCAdynamics}. a) Intermediate scattering function $F(q,t)$ ($q=2\pi/\sigma_L$) as a function of time at low temperature $k_B T / \epsilon = 10^{-4}$, and at several densities as indicated. HS systems at the equivalent effective density (mapped using Eq. \ref{eq:effHSfit}) are shown as dotted lines. b) Relaxation time $\tau_{\alpha}$ as a function of the wave-vector $q \sigma$ for multiple temperatures, at density $\rho \sigma^3_L = 1.175$. Error bars are determined as twice the standard error.
    }
    \label{figure:biWCAdynamics}
\end{figure}

For the binary WCA mixture, we first compare the dynamics obtained from EDMC and MD in Fig. \ref{figure:biWCAdynamics}.
In Fig. \ref{figure:biWCAdynamics}a we show the ISF $F(q,t)$ at low temperature for different densities, which in contrast to the monodisperse case shows the clear formation of a plateau at intermediate times.
This plateau, which lengthens with increasing system density, can be attributed to the caging of particles by their neighbors, and is  typical of glassy systems.
We notice a much larger range of relaxation times compared to the monodisperse case (figure \ref{figure:biWCAdynamics}b) which implies much longer simulation times are required, even for the sole purpose of equilibration. Again, EDMC accurately reproduces the dynamics of conventional MD, with the LAMMPS simulations possibly having a slightly shorter relaxation time, although this difference is hard to resolve accurately. Similarly to the monodisperse case, we also recover the expected relaxation dynamics over a large span of temperatures and length scales, as shown in figure \ref{figure:biWCAdynamics}b.

\begin{figure}
    \raggedright
    a)\\
    \centering
    \includegraphics[width=\columnwidth]{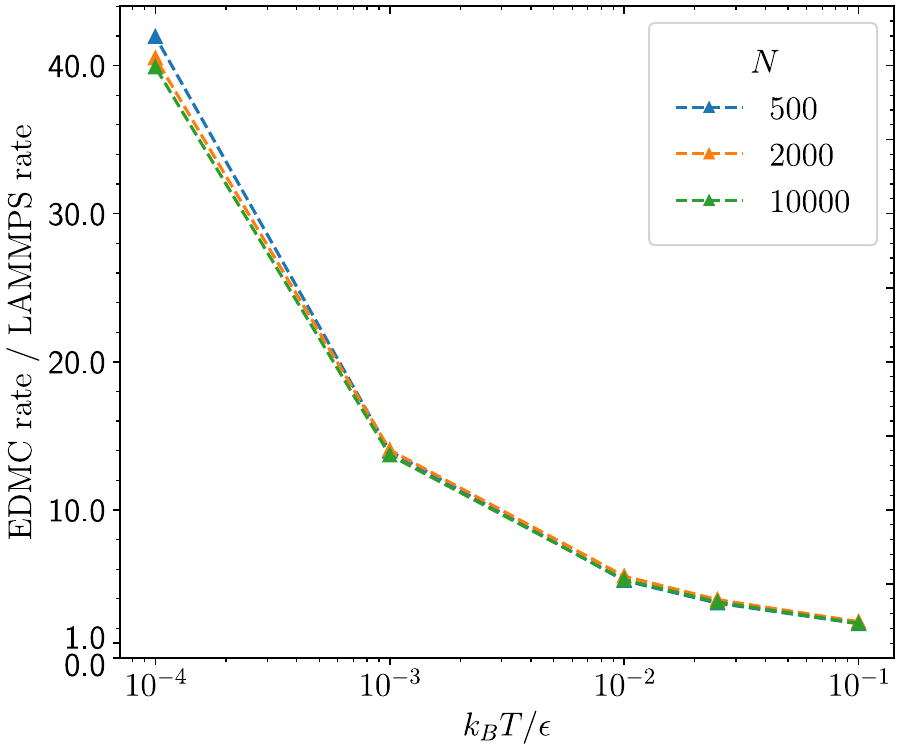}\\
    \raggedright
    b)\\
    \centering
    \includegraphics[width=\columnwidth]{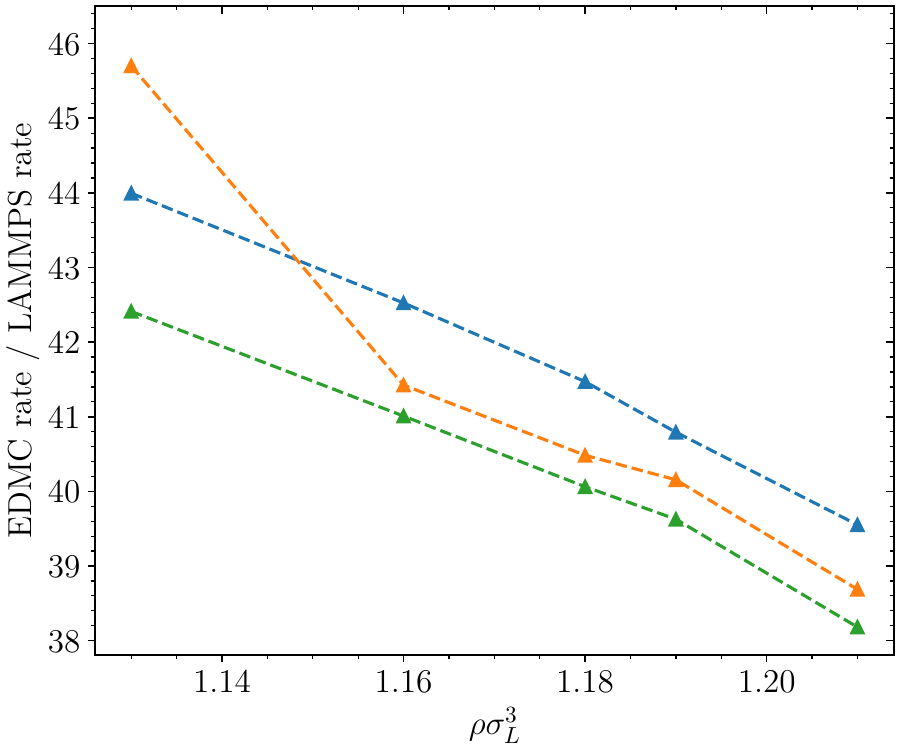}
    \caption{
    Relative performance of the EDMC and MD methods for several system sizes, expressed as the ratio of the number of time units simulated per second in each simulation method, a) at fixed density $\rho \sigma_L^3 = 1.175$ and b) at fixed temperature $k_B T / \epsilon = 10^{-4}$. Performance evaluation was obtained from successive short simulations of duration $100 \tau_{kT}$ started from an equilibrated fluid system. Standard deviations are typically found to be much less than $1\%$.
    }
    \label{figure:biWCAbench}
\end{figure}

In Fig. \ref{figure:biWCAbench}, we compare the simulation speed of the EDMC method to the LAMMPS package as a function of temperature (Fig. \ref{figure:biWCAbench}a) and density (Fig. \ref{figure:biWCAbench}b). In particular at low temperatures, we again find a significant speedup (up to a factor of approximately 40).

The efficiency of the EDMC method at low temperatures allows us to use this simulation approach to look at the interplay between glassy dynamics and Laves phase nucleation at temperatures lower than those that were easily accessible to the time-driven molecular dynamics methods previously used to study nucleation in this system (Ref. \onlinecite{dasgupta2020tuning}). To this end, we run brute-force nucleation simulations for a range of densities $\rho \sigma^3$ and temperatures $k_B T/\epsilon = 0.005$ and  $k_B T/\epsilon = 10^{-4}$, using system sizes of $N=2000$ particles. We run our simulations for at least $6 \times 10^6 \tau_\textrm{kT}$. For comparison, the lowest temperature examined in Ref. \onlinecite{dasgupta2020tuning} was $k_B T/\epsilon = 0.005$.

\begin{figure}
    \centering
    \includegraphics[width=\columnwidth]{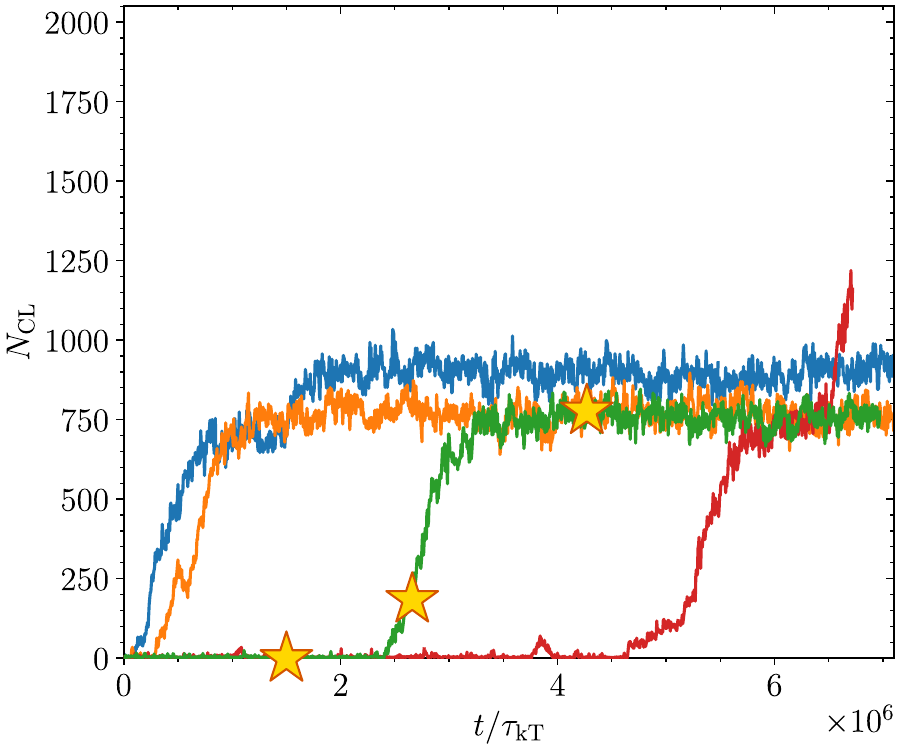}\\
    \centering
    \includegraphics[width=0.325\columnwidth]{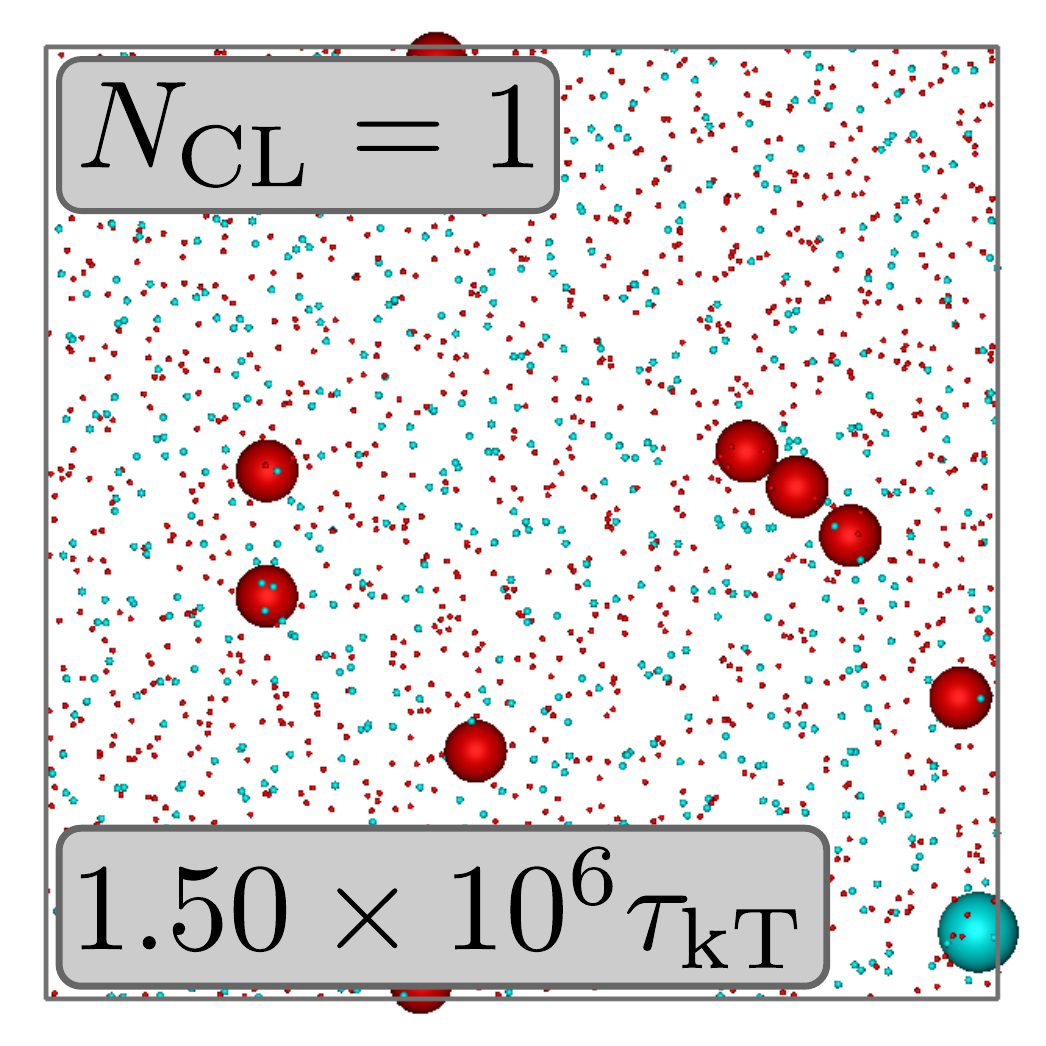}
    \includegraphics[width=0.325\columnwidth]{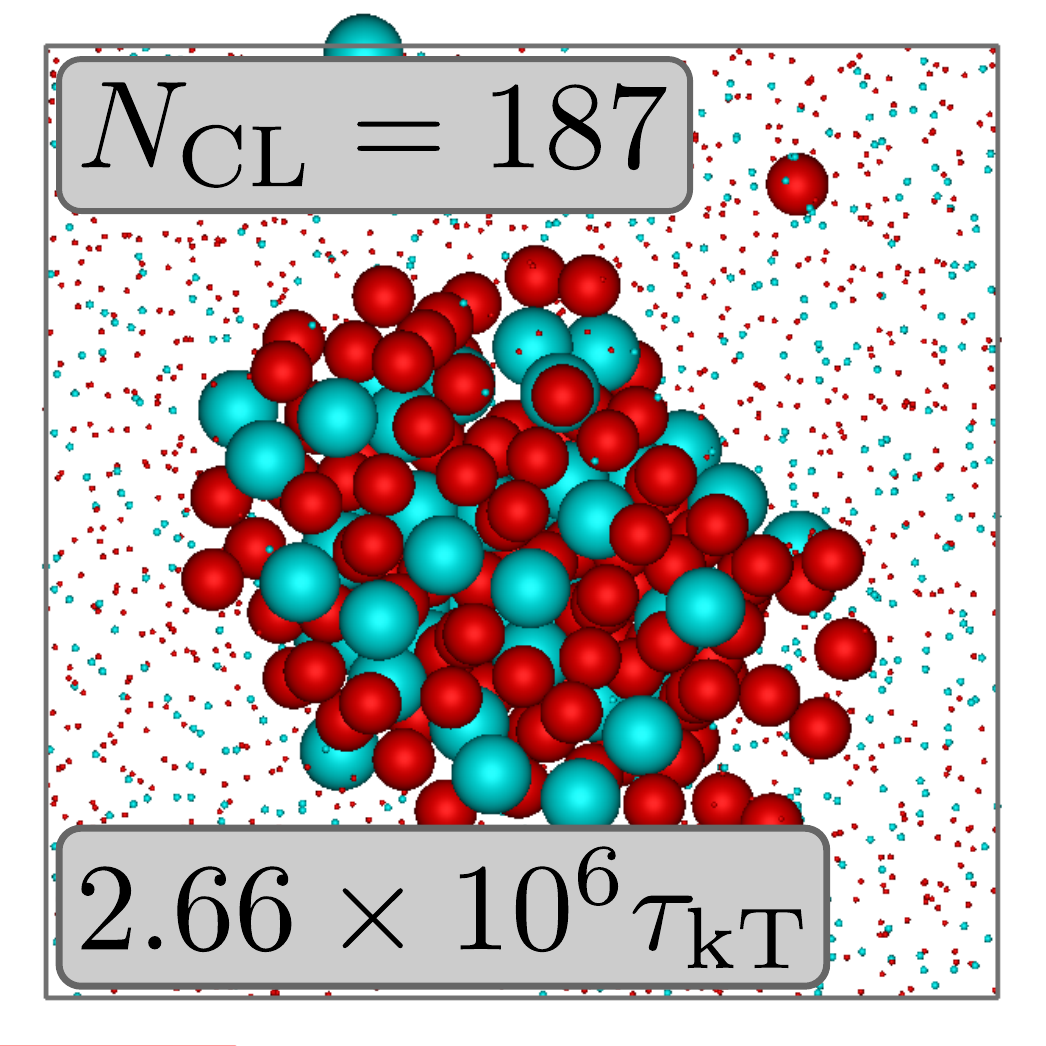}
    \includegraphics[width=0.325\columnwidth]{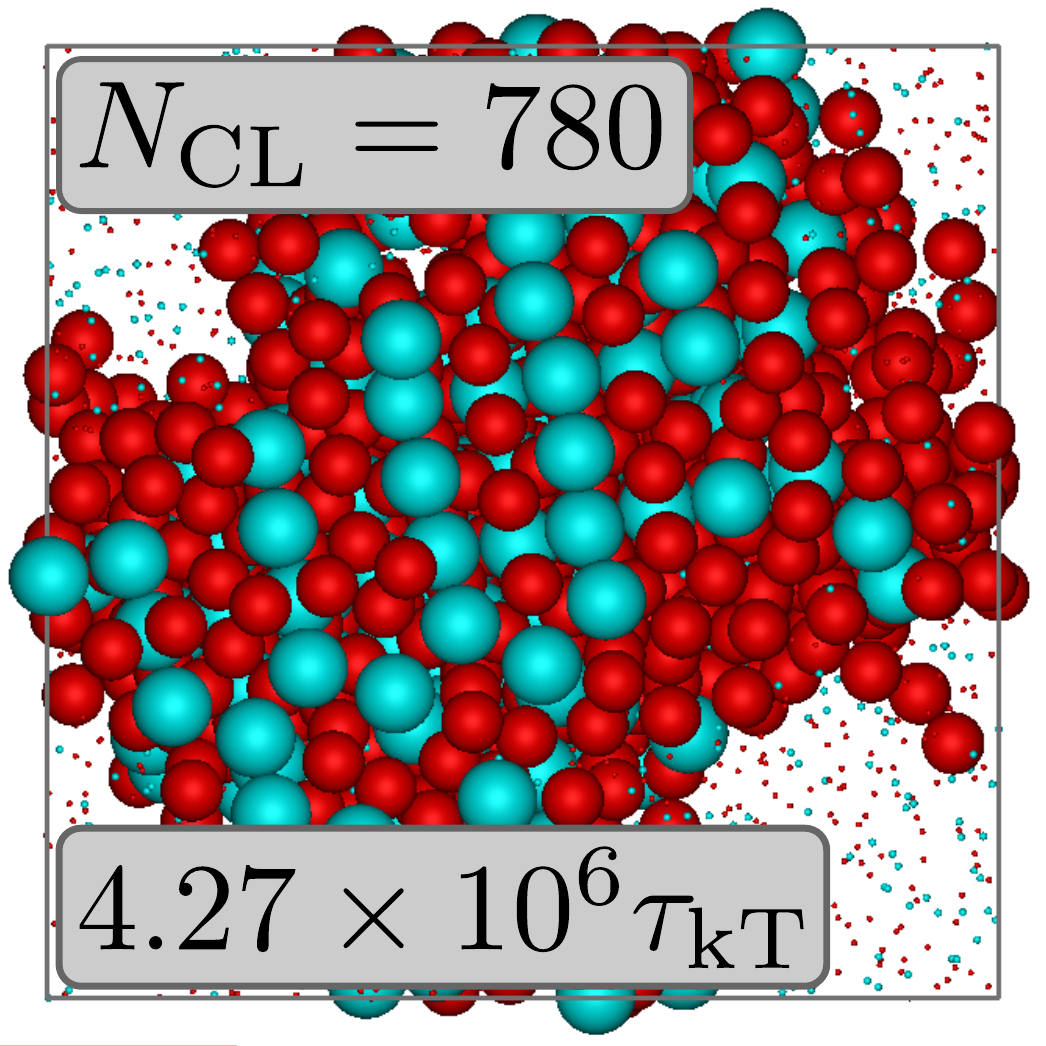}
    \caption{Nucleation events at low temperature for the binary WCA system ($x_\mathrm{L} = 1/3$, $\gamma = 0.78$), showing multiple nucleation events with waiting time at the same state point ($k_B T / \epsilon = 0.005$ , $\rho \sigma_L^3 = 1.213$). The total system consisted of $N=2000$ particles. The snapshots show the evolution of one system (green line) undergoing crystallization. Large particles are depicted in blue and small particles in red. The yellow stars in the plot report where each snapshot of the system was taken.
    }
    \label{figure:biWCAnucleation}
\end{figure}

In contrast to Ref. \onlinecite{dasgupta2020tuning}, we do observe spontaneous nucleation at a temperature $k_B T/\epsilon = 0.005$. As an example of this, we show in Fig. 
\ref{figure:biWCAnucleation} the growth of the largest connected cluster of particles in a Laves-phase-like environment (see Methods). In particular, we show four nucleation trajectories taken at the same state point inside the coexistence region ($\rho\sigma^3 = 1.213$, $k_BT/\epsilon = 0.005$). We observe a clear (random) waiting time followed by a jump to a finite-sized cluster consisting of approximately half of the total number of particles in the system ($N = 2000$). Analysis of the system using our order parameter shows that this state corresponds to a roughly spherical cluster (second snapshot in Fig. \ref{figure:biWCAnucleation}). Running the simulation longer, we also sometimes observe a second transition to a system-spanning cluster (third snapshot in Fig. \ref{figure:biWCAnucleation}). This can sometimes be seen in the evolution of the cluster size over time as a secondary jump to a larger value (as visible in the red line in Fig. \ref{figure:biWCAnucleation}).
In terms of structure composition, we typically observe a mixture of the competing MgZn$_2$ and MgCu$_2$ Laves phases, with MgNi$_2$ appearing only in small amounts, and typically at later stages of the simulation. Similar to what is reported in Ref. \onlinecite{dasgupta2020tuning}, it appears that MgZn$_2$-like environments emerge rapidly from the metastable fluid and that they later transform into MgCu$_2$-like environments as the crystal nucleus extends across the simulation box. These observations are consistent with the previously reported small ($\sim 10^{-3} k_B T$ per particle) free-energy difference that exists between these phases \cite{hynninen2009stability}.

\begin{figure}
    \centering
    \includegraphics[width=0.48\textwidth]{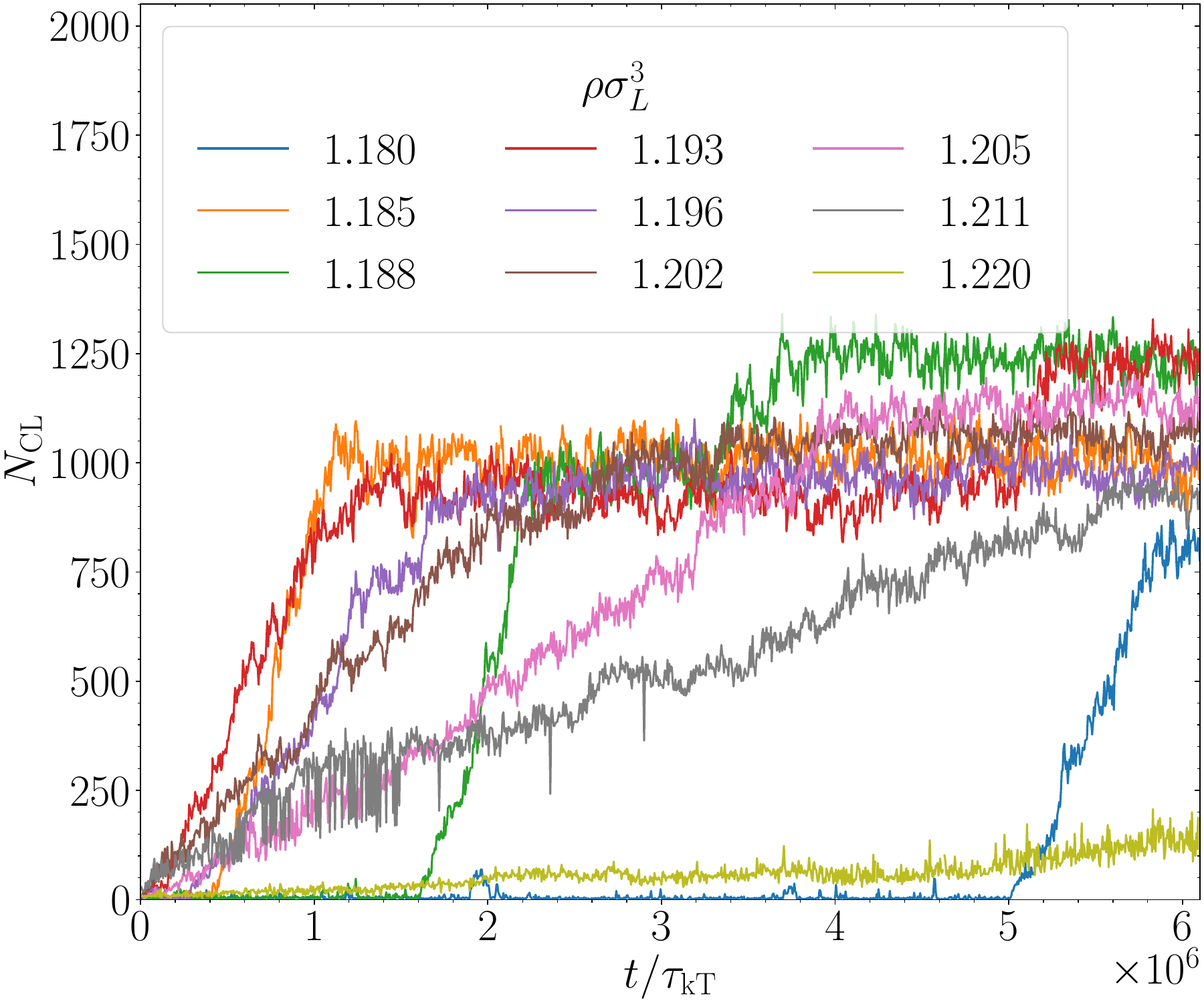}
    \caption{Nucleation events recorded at very low temperature ($k_B T / \epsilon = 10^{-4}$) for a binary WCA system with composition $x_\mathrm{L} = 1/3$ and size ratio $\gamma = 0.78$ for a range of densities.}
    \label{figure:biWCAnuc_lowT}
\end{figure}

We note that the state point investigated in Fig. \ref{figure:biWCAnucleation} lies below the ``instability line'' determined by Dasgupta \textit{et al.}\cite{dasgupta2020tuning}, which they defined as the temperature below which the supersaturated fluid becomes unstable to crystallization and hence where they expected no discrete nucleation events to be observable.
We also report in Fig. \ref{figure:biWCAnuc_lowT} a selection of nucleation trajectories recorded at an even lower temperature $k_B T/\epsilon = 10^{-4}$, at different densities. As one might expect, we observe a random waiting time at low densities ($\rho \sigma^3_L \lesssim 1.188$), and an immediate onset of crystallization at higher densities. On further increasing the density, crystallization still sets in immediately, but the crystal growth becomes slower, due to the overall slowdown of the dynamics. In these glassy systems, we also sometimes see jumps in the cluster size that rapidly revert (see e.g. the curve associated with $\rho\sigma^3_L = 1.211$). We attribute this to the crystal nucleus consisting of multiple domains that have joined, but are not fully aligned yet, such that thermal fluctuations may temporarily `break' the connection between parts of the cluster. In our simulations at this temperature, we no longer observe crystallization for densities higher than $\rho \sigma^3_L \gtrsim 1.22$, as the system becomes too glassy. A broader selection of nucleation trajectories can be found in the Supplementary Material (SM). 

In contrast to the findings of Ref. \onlinecite{dasgupta2020tuning}, we do not find a low-temperature regime where spontaneous crystal nucleation and growth is completely pre-empted by glassy behavior. This may be attributable to the ability of the EDMC approach to reach longer time scales at low temperatures than the conventional MD simulations of Ref. \onlinecite{dasgupta2020tuning}. Additionally, Dasgupta \textit{et al.} used Monte Carlo (MC) simulations to determine the glass transition for their system. As MC simulations have their own microscopic dynamics (which can depend on the chosen step size), this could influence the predicted critical packing fraction associated with dynamical arrest. Note that our observation of nucleation and growth of the Laves phases at low temperatures is consistent with past observations of Laves phase nucleation in pure hard spheres at similar size ratios \cite{bommineni2020spontaneous, marin2020tetrahedrality}, which can be seen as the low-temperature limit of the WCA model considered here.

\section{Conclusions}
\label{section:conclusion}

We have demonstrated that the rejection-free event-driven Monte Carlo approach introduced by Peters and de With \cite{peters2012rejection}, combined with an efficient implementation of event-driven molecular dynamics \cite{smallenburg2022efficient}, provides an accurate and efficient method for simulating the dynamics of model systems with sharply varying interactions like the WCA model. At sufficiently low temperatures, this approach circumvents the time-step issues encountered by molecular dynamics simulations, resulting in significant boosts in speed.

We should acknowledge here that we have considered an ideal case for the EDMC method: a short-ranged, purely repulsive interaction potential, which can be easily analytically inverted to obtain an expression for the distance as a function of interaction energy. 
When considering other systems, the suitability of EDMC as a substitute for MD simulations comes with a number of caveats.

First, every collision prediction in an EDMC simulation requires the evaluation of both $U_{ij}(r)$ and its inverse $r(U_{ij})$. This means that the function should ideally be analytically invertible. In turn, this would incur a significant performance penalty if one were to solve numerically for the inverse during every collision. However, this issue could in principle be addressed with e.g. lookup tables, or by avoiding most exact evaluations of the potential using an approximate inverse of the potential (thinning) \cite{tartero2024concepts}.

Second, for interaction potentials with both attractive and repulsive components (e.g. Lennard-Jones), collision prediction needs to check for both attractive and repulsive collisions \cite{peters2012rejection}, similar to EDMD simulations of square-well systems. This will slow down the simulation speed. Note that oscillatory interaction potentials (with multiple switches between attractive and repulsive regimes) would exacerbate this slowdown. Preliminary tests on the Lennard-Jones model show that the EDMC model can still be competitive with conventional MD. 

Third, as potentials become more long-ranged, the size of the neighbor lists increases. In event-driven simulations of large systems, this will significantly increases the cost of collision calculations due to scattered memory access.

Fourth, we note that similar to EDMD, EDMC could in principle be adapted to non-spherical interactions. However, for the vast majority of anisotropic models, an analytical prediction of a collision time is not feasible. Numerically solving for collision times (see e.g. \cite{hernandez2007discontinuous}) would be significantly slower, and unlikely to be competitive with conventional MD.

Fifth, the EDMC approach inherently simulates a constant-temperature ensemble. While this may be advantageous in some cases, there are scenarios where energy conservation is crucial, e.g. when studying heat transport through a system. 

Finally, despite a few existing routes in this direction\cite{miller2004event, khan2011parallel}, we point out that EDMC, similar to EDMD, is difficult to parallelize efficiently, which may limit its suitability to extremely large-scale simulations.

The EDMC approach also offers several advantages (apart from its efficiency under suitable conditions).
Most important is the ability of EDMC to simulate arbitrarily steep interaction potentials, without impacting the required time step. Second, unlike either MD or EDMD, EDMC can handle interactions which combine discontinuous and continuous elements. Examples of these would be e.g. hard-spheres combined with a screened-Coulomb repulsion, or an Asakura-Oosawa depletion attraction.
Third, the avoidance of a finite integration step size also implies that EDMC circumvents any systematic inaccuracies this discretization might introduce: the EDMC approach simulates the implemented interaction potential exactly, without requiring the approximations necessary for MD.
Finally, in systems that are (at least partially) in a gas phase, EDMC will be highly efficient in comparison to MD, as the free flight of particles is extremely computationally cheap in event-driven simulations.

In conclusion, for studying the dynamics of suitable model systems, the EDMC approach is an appealing alternative to molecular dynamics simulations in the canonical ensemble, with potential speedups of over an order of magnitude.

\section{Supplementary Material}

Additional details on random number generation as well as additional low temperature nucleation trajectories may be found in the Supplementary Material available online.

\section{Acknowledgments}

AC and FS gratefully acknowledge funding from the Agence Nationale de la Recherche (ANR), grant ANR-21-CE30-0051. LF acknowledges funding from the Vidi research program with project number VI.VIDI.192.102 which is financed by the Dutch Research Council (NWO).

\section{Author Declarations}

\subsection{Conflict of Interest}

The authors have no conflicts to expose.

\section{Data Availability}

A version of the EDMC simulation code can be found at \href{https://github.com/FSmallenburg/EDMC/}{https://github.com/FSmallenburg/EDMC/}. 
The data that support the findings of this study is available in a data package on Zenodo \cite{castagnede2024EDMCdatapackage}.

\bibliography{refs}

\end{document}